\documentstyle[11pt,newpasp,twoside,epsf]{article}
\def\edcomment#1{\iffalse\marginpar{\raggedright\sl#1\/}\else\relax\fi}
\reversemarginpar

\begin{document}
\title{Multiwavelength Aspects of Stellar Coronae}
\author{Manuel G\"udel}
\affil{Paul Scherrer Institut, W\"urenlingen \& Villigen, CH-5232 Villigen PSI, Switzerland}

\begin{abstract}
Coronae express different facets of their energy release processes 
in different wavelength regions. While soft X-ray and EUV emission
dominates the radiative losses of the thermal plasma,  
hard X-ray emission ($>$10 keV) can be produced from  non-thermal high-energy 
particles accelerated as a consequence of flare processes, and 
radio emission is often emitted from mildly relativistic electrons trapped
in magnetic fields. Combined measurements of all emissions are important
to understand the ultimate mechanisms responsible for coronal energy
release and heating. This presentation will summarize a number of
aspects for which multi-wavelength studies are important, both for 
quiescent and flare emission.
\end{abstract}

\section{Introduction}  

It has become practice to think of stellar coronae as - perhaps more
luminous - equivalents
to the wonderfully (and colorfully) shining outer {\it solar} atmosphere
on Skylab, Yohkoh, SoHO, and TRACE images. The vast progress in the
areas of solar EUV and X-ray astronomy, in particular of imaging, 
and the immense flood of {\it stellar} data from a completely new X-ray 
discipline - that of high-resolution stellar coronal X-ray spectroscopy 
from {\it XMM-Newton} and {\it Chandra} - more than justifies a place of 
honor for this view. But don't we miss a point? Do soft X-rays 
provide a self-contained description of a stellar corona, of its 
energetics, of its structure, of the physics at work, of the essentials
that make up `the corona'?

The present review tries to illuminate aspects of stellar
coronal physics from alternative angles, in particular related
to non-thermal phenomena and their relation to the soft X-ray radiating
plasma. The main goal is to provide contrasting and complementary points
to a conference largely devoted to new X-ray results. The theme 
focuses on the important role that high-energy processes play
in a corona - invisible to X-ray detectors, often occurring before
the appearance of X-ray signatures, but very relevant to the `big picture'
of what the term `corona' really stands for.

\section{Why Multi-Wavelength Coronae?}  

During flaring episodes, the solar corona becomes a truly 
multi-wavelength object. Often before any significant soft X-ray brightening
occurs, tangled magnetic fields produce immense numbers of high-energy
particles up to relativistic energies. Evidence for non-thermal particles 
is ubiquitous; each solar flare seems to produce them, in a wide variety 
(deka-keV electrons, relativistic electrons, high-energy ions, high-energy 
neutrons, pions), and they contain a significant fraction of the total 
energy released in a flare (Hudson \& Ryan 1995). If their power-law distribution 
in energy is not rigidly cut off toward lower ($\sim 10-20$~keV) energies, 
embarrassingly large total energies could be contained in this population 
(Dennis 1988). The production of high-energy particles is an efficient way to 
release energy built up in non-potential magnetic fields in tenuous plasma environments, 
and they are efficient agents to transport this energy to different locations 
in very short times. Upon collisions in denser environments, they produce
hard X-rays ($>10$ keV) and $\gamma$ rays. Mildly relativistic electrons
trapped in magnetic fields emit radio gyrosynchrotron emission. The short
time scales involved (seconds, compared to minutes for soft X-ray plasma)
are ideal tracers of processes occuring at the time of the initial
energy release.

It was an unanticipated but momentous discovery that magnetically active 
stars are quasi-steady emitters of {\it non-thermal} gyrosynchrotron emission
(e.g., review by Dulk 1985). It testifies to the presence of a large number
of relativistic electrons in stellar coronae. How are they accelerated, and
where does this particle population carry its energy to? The wide-ranging
analogy with solar flares provokes further questions: 
i) Is there also a stellar non-thermal hard X-ray component? 
ii) Are the high-energy particles responsible for coronal heating? 
iii) Is the apparently quiescent emission made up of flares?

Non-thermal {\it stellar} hard X-ray emission has not yet been detected given the
instrumental limitations. But thanks to detailed studies at radio
wavelengths, the initial phase of stellar flares and energy release 
can be studied, with potential implications for our understanding 
of coronal heating, and {\it spatially resolved} images of stellar
coronae (only one aspect of them, alas!) can be taken.

\section{Coronal Structure}  

The spatial structure of non-thermal coronae is routinely accessible
to Very Long Baseline Interferometry (VLBI) at  milliarcsec scales (Table 1).
It was soon recognized that RS CVn and Algol-type binaries are surrounded
by large, global-scale non-thermal coronae (Mutel et al. 1985), often
composed of a compact core plus an extended halo. The former is thought
to relate to initial phases of flares, while the latter contains 
slowly decaying high-energy particles injected from the flare site. 

Similar spatially-resolved observations have been obtained from T Tau stars
(Phillips, Lonsdale, \& Feigelson 1991) and magnetic chemically peculiar Bp/Ap stars
(Phillips \& Lestrade 1988, Andr\'e et al. 1991). The have mostly been interpreted
in terms of global, dipole-like magnetic fields surrounding the active
star(s) on spatial scales exceeding the stellar size by far. In the most
elaborate models, the magnetic fields are compressed (e.g., by a weak stellar 
wind escaping at high latitudes) toward the equatorial plane where they
build up current sheets that accelerate electrons. These are trapped within 
the magnetospheres, somewhat similar to the Earth's van Allen belts (Drake et
al. 1987, Andr\'e et al. 1988, Morris, Mutel \& Su 1990, Linsky, Drake,
\& Bastian 1992).  
 
VLBI techniques have been much more demanding for single late-type dwarf stars, both due
to lower flux levels and (supposedly) smaller coronal sizes. Some  observations
with mas resolutions show unresolved (quiescent and flare) sources, thus 
providing stringent constraints on the radio brightness temperature ($T_b > 10^{10}$~K)
and therefore on the (non-thermal) energies of the involved electrons (Benz \& Alef 1991, Benz,
Alef, \& G\"udel 1995), while others
have shown evidence for extended coronae, such as for the eclipsing YY Gem, with
coronal sizes up to several times the stellar size (Alef, Benz, \& G\"udel 
1997, Pestalozzi et al. 1999).

\begin{figure}[t!]
\plottwo{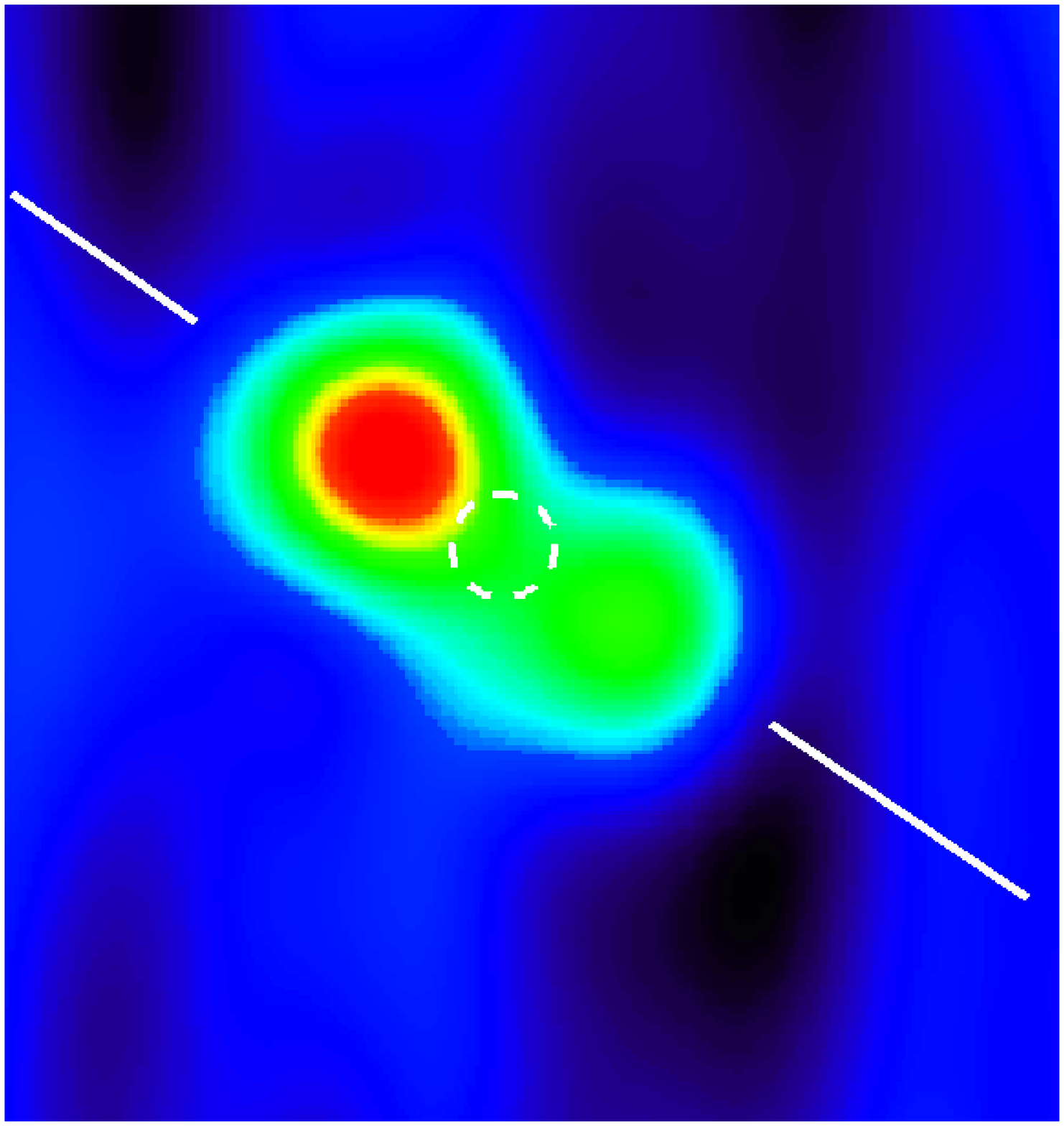}{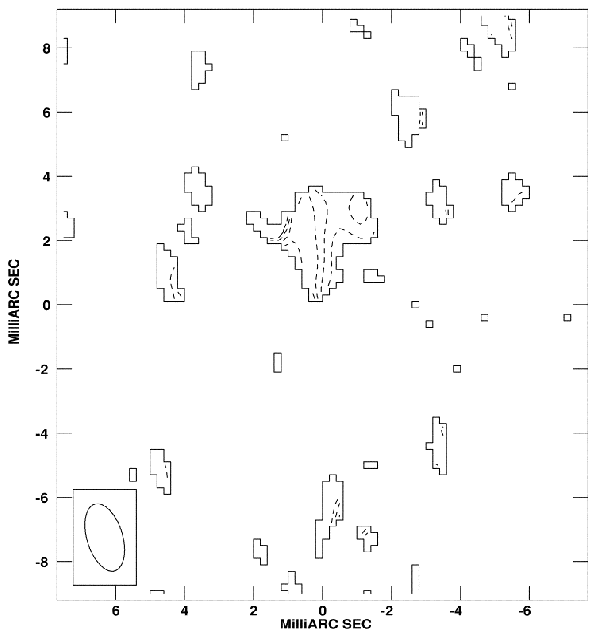}
\caption{{\bf Left:} The dMe star UV Cet B as seen at radio wavelengths by the VLBA.
The separation between the two components is about 1.4~mas, while the best angular 
resolution is 0.7~mas. The circle
indicates the photospheric diameter of UV Cet B, and the solid line marks the
putative rotation axis - assumed to be parallel to the axis of the visual-binary orbit
(after Benz et al. 1998). - {\bf Right:} Radio polarization map of the spatially resolved
RS CVn binary UX Ari. Note the gradient in the polarization contours 
across the stellar image (from Beasley \& G\"udel 2000).}
\end{figure} 

The dMe star UV Cet B reveals a pair of giant synchrotron lobes (Figure 1), with sizes up to 
$2.4\times 10^{10}$~cm and separated by 4$-$5 stellar radii along the  putative 
rotation axis of the star, suggesting very extended magnetic structures above the 
magnetic poles (Benz, Conway, \& G\"udel 1998). An analysis of the 
synchrotron losses constrains the strength of extended magnetic fields 
to $15-130$~G. VLBA imaging and polarimetry of Algol 
reveals a similar picture (Mutel et al. 1998) with two oppositely polarized radio lobes 
separated perpendicularly to the orbital plane by more than a stellar diameter of the 
K star. Global polarization structure has been found in UX Ari, indicating  the presence 
of large-scale ordered magnetic fields on size scales of the intrabinary
distance (Beasley \& G\"udel 2000).

\begin{table}[t!]
\caption{SELECTED  VLBI RESULTS FOR ACTIVE STARS}
\begin{tabular}{lllll}
               &                      &                           &              &       \\
\hline
\hline
Star           &      $R_*$~$^a$          & $R_{\mathrm{c}}/R_*$~$^b$    &  model$^c$   & Reference  \\
               &      (cm)           &                            &          &    \\
\hline
\multicolumn{5}{l}{\it RS CVn and Algol Binaries}    \\
$\sigma$ CrB   & $8.4\times 10^{10}$ & 1.2                        &          & Mutel et al. 1985 \\
HR~1099        & $2.7\times 10^{11}$ & 1.6                        &          & Mutel et al. 1985 \\
HR~5110        & $2.0\times 10^{11}$ & 1.7                        &          & Mutel et al. 1985 \\
UX Ari         & $3.3\times 10^{11}$ & 3.6                        &   halo   & Mutel et al. 1985 \\
UX Ari         & $3.3\times 10^{11}$ & $<$0.45                    &   core   & Mutel et al. 1985 \\
Algol          & $2.4\times 10^{11}$ & 2.5                        &   halo   & Lestrade et al. 1988 \\
Algol          & $2.4\times 10^{11}$ & $<$0.5                     &   core   & Lestrade et al. 1988 \\
Algol          & $2.4\times 10^{11}$ & 1.4                        &   polar lobes & Mutel et al. 1998 \\
UX Ari         & $3.3\times 10^{11}$ & 4.5                        &   MS     & Beasley \& G\"udel 2000 \\
\multicolumn{5}{l}{\it Pre-Main Sequence Stars}    \\
HD~283447      & $2.1\times 10^{11}$ & 12$-$15                    &   MS     & Phillips et al. 1991 \\ 
HDE~283572     & $2.4\times 10^{11}$ & $<$14                      &   MS     & Phillips et al. 1991 \\ 
Hubble 4       & $2.4\times 10^{11}$ & 12$-$24                    &   MS     & Phillips et al. 1991 \\ 
DoAr 21        & $2.4\times 10^{11}$ & 10$-$12                    &   MS     & Phillips et al. 1991 \\ 
\multicolumn{5}{l}{\it Chemically Peculiar Stars }    \\
$\rho$ Oph S1  & $3.0\times 10^{11}$ & 6.4$^{+1.2}_{-2.4}$        &   MS             & Andr\'e et al. 1991 \\ 
$\sigma$ Ori E & $2.8\times 10^{11}$ & $\le$6                     &   MS             & Phillips \& Lestrade 1988 \\ 
\multicolumn{5}{l}{\it Main-Sequence Stars }    \\
AD Leo         & $2.0\times 10^{10}$ & $<$1.8                     &                  & Pestalozzi et al. 1999 \\
AD Leo         & $2.0\times 10^{10}$ & $<$1.9                     &                  & Benz et al. 1995 \\
AD Leo         & $2.0\times 10^{10}$ & $<$3.7                     &                  & Benz et al. 1995 \\
YZ CMi         & $2.4\times 10^{10}$ & 1.7$\pm$0.3                &                  & Pestalozzi et al. 1999 \\
YZ CMi         & $2.4\times 10^{10}$ & $<$1.7                     &                  & Benz \& Alef 1991 \\
EQ Peg         & $1.3\times 10^{10}$ & $<$1                       &                  & Benz et al. 1995 \\
YY Gem         & $4.8\times 10^{10}$ & 2.1$\pm$0.6                &                  & Alef et al. 1997 \\
UV Cet         & $1.0\times 10^{10}$ & 2.2$-$4                    &    polar lobes   & Benz et al. 1998 \\
\hline
\hline
\multicolumn{5}{l}{Notes: $^a$ Information used from Hipparcos Catalog (ESA 1997) and}\\
\multicolumn{5}{l}{Strassmeier et al. 1993}\\
\multicolumn{5}{l}{$^b$ $R_{\mathrm{c}}$ = coronal radius, $R_*$ = stellar photospheric radius. }\\
\multicolumn{5}{l}{$^c$ proposed geometric strucutre; MS = magnetospheric}  \\  
\end{tabular}
\end{table}

Some coronal structure information can also be derived from rotational 
modulation and eclipses. The imaged large-scale magnetospheric structures would 
suggest that this approach is less promising, and there are indeed not many
clear-cut examples available. Some optically eclipsing systems such as YY Gem
or AR Lac simply show no radio eclipses (Alef et al. 1997, Doiron \& Mutel 1984),
suggesting that the coronal structures are very large or completely 
outside the eclipse zone. Some light-curve modeling even suggests the 
presence of radio-emitting material between the binary components of 
such systems (Gunn et al. 1997, 1999).

Are these structures co-spatial with the X-ray coronae implied for these stars? 
Pressure balance for the hot thermal X-ray plasma would imply electron densities 
$n_{\mathrm{e}}$ less than a few times $10^8$~cm$^{-3}$ (Beasley \& G\"udel 2000); 
the requirement that the local plasma frequency $\nu_{\mathrm{p}} = 
(n_{\mathrm{e}}e^2/(\pi m_{\mathrm{e}}))^{1/2}$ be less than the observing frequency 
($\sim$1~GHz) restricts  $n_{\mathrm{e}} < 10^{10}$~cm$^{-3}$. Both values contradict
explicit X-ray density measurements (G\"udel et al. 1999, 2001ab). The low densities
are also not compatible with the rapid decay time scales usually seen in stellar 
X-ray flares. Further, the high-energy electrons in a termal-plasma environment
thermalize on time-scales of $\tau = 1.6\times 10^{12}E_{\mathrm{MeV}}/n_{\mathrm{e}}
\approx 1-100$~s (for likely parameters; Benz \& Gold 1971). Lastly, free-free absorption 
along the line of sight is $\propto n_{\mathrm{e}}^2\ell$ with $\ell$ being the 
characteristic source depth - again, the
opacity would suppress most of the emission. The deep X-ray 
eclipse of the YY Gem binary restricts the X-ray  emitting plasma to intermediate 
and low latitudes and the scale height to $\le 1R_*$ (G\"udel et al. 2001a), at 
variance with the VLBI measurements. These arguments make it unlikely that 
the large radio structures are spatially related to the X-ray coronae.

\section{Energy Release: Evidence from Solar Flares}  

Are radio and X-ray coronae linked energetically, in particular during flares?
We first review some important evidence from the Sun. Solar flare hard X-ray emission
correlates with microwave radiation both in peak flux and,
to sub-second accuracy, in time (e.g., Kosugi, Dennis, \& Kai 1988). Such correlations 
are now taken as evidence that both the $\sim$10$-$100~keV electrons responsible for hard X-rays
and the higher-energy radio emitting electrons belong to one and the same non-thermal
parent population.

A standard flare scenario devised from many solar observations proposes
that accelerated coronal electrons precipitate
into the chromosphere where they lose their kinetic energy by collisions, thereby
heating the cool plasma to coronal flare temperatures. The subsequent overpressure
drives the hot material into the coronal loops, giving rise to a soft X-ray flare.
The radio gyrosynchrotron emission from the accelerated electrons is roughly proportional 
to the instantaneous number of particles  and therefore to the power injected into the 
system. On the other hand, the X-ray luminosity is roughly proportional to the total energy
accumulated in the hot plasma. One thus expects, to first order,
\begin{equation} 
L_R(t) \propto {d\over dt}L_{\rm X}(t)
\end{equation}
which is known as the ``Neupert Effect'' (Neupert 1968) and has been well observed
on the Sun in most impulsive and many gradual flares (Dennis \& Zarro 1993). Further 
evidence for the operation of the chromospheric evaporation scenario include:
\begin{itemize}
\item Footpoint brightening of distant ($\sim 10^9$~cm) loop footpoints within
  a fraction of a second, involving non-thermal speeds for the energy transport
  (Sako 1994, Hudson \& Ryan 1995);
  
\item Prompt soft X-ray footpoint brightening together with hard X-rays, while the
   bulk volume of the extended loop follows later, according to equation (1)
   (Hudson et al. 1994);
   
\item Spatially related and temporally correlated  evolution of soft X-ray, hard
     X-ray, and radio structures (Nishio et al. 1997);
     
\item Correlated hard X-ray and white light emission, both as a consequence of 
     impacting fast electrons (Hudson et al. 1992);
     
\item Blueshifted Ca~XIX enhancements that are closely correlated with hard X-rays, signifying
     the rapidly developing upflows during the impact episode (Bentley et al. 1994).

\end{itemize}

\section{Energy Release and Radio Flares in Stars}  

The search for stellar equivalents to the solar Neupert effect 
has been a story of contradictions if not desperation.
A first breakthrough came with the simultaneous EUV (a proxy for X-rays) and optical 
observations (a proxy for the radio emission) of a flare on AD Leo (Hawley et 
al. 1995) and radio + X-ray observations of several flares on UV Cet (G\"udel et al. 1996).
The relative timing between the radio bursts and the gradual X-ray flares was found to
be  similar to solar equivalents (Figure 2a), including the energy ratios seen in 
non-thermal and thermal emissions. These observations gave conclusive evidence
that the production of high-energy particles accompanies flare coronal heating, 
and may actually be the cause for the heating itself through chromospheric evaporation.

Given the large size of RS CVn radio magnetospheres, are there similar physical
mechanisms at work in such coronae as well? It is conceivable that radio-emitting
particles are more detached from the heating site and may not be responsible for
chromospheric evaporation. A recent observation with {\it XMM-Newton} and the
VLA seems to indicate otherwise. Figure 2b shows, from top to bottom, the
X-ray light curve, its time derivative, and the radio light curve during a large
flare episode on the RS CVn binary $\sigma$ Gem. We concentrate on the second flare that is fully
visible at radio wavelengths, i.e., between 1.02$-$1.32~d.
The radio and X-ray derivative curves correlate very well
in time,  with no significant relative time delay. Evidently, the release of high-energy particles
is closely coupled with the heating mechanism. Correlated behavior does, however,
not prove the operation of chromospheric evaporation as both mechanisms may be
unrelated consequences of the energy release process. A necessary condition is
that the energetic particles carry enough kinetic energy to explain the
released soft X-ray energy. Under simplified assumptions such as an 
electron power-law distribution in energy, a reasonable lower energy cutoff around
10~keV, and the absence of strong changes in the radio optical depth, an order of
magnitude estimate of the total energy in the electrons results in $E_{\mathrm{tot}}
= 10^{33} - 10^{36}$~erg (G\"udel et al. 2001c). This estimate holds for a magnetic
field strength $B$ between 20$-$200~G and a power-law index $\delta =2.0-3.5$, values
that have previously been found from magnetospheric modeling, and an
electron lifetime of $\sim$1500~s as inferred from the light curve (shorter 
lifetimes result in larger energies). The total released energy in the X-ray flare
is estimated to be $4\times 10^{34}$~erg. The kinetic particle energy is thus
sufficient to explain the heating for a broad range of parameters $B, \delta$.
 
In retrospect, we find a similar 
timing between radio and X-ray flare events in some previously published light curves, although 
the Neupert effect was not discussed. Most evidently, radio emission peaking before the soft 
X-rays, thus suggesting the presence of a Neupert effect,  can be seen in the examples presented by
Vilhu et al. (1988),  Stern et al. (1992), and Brown et al. (1998).
To conclude this section, it is important to note that the Neupert effect is neither 
observed in each solar  flare (Dennis \& Zarro 1993) nor in each stellar flare.

\begin{figure}[t!]
\plottwo{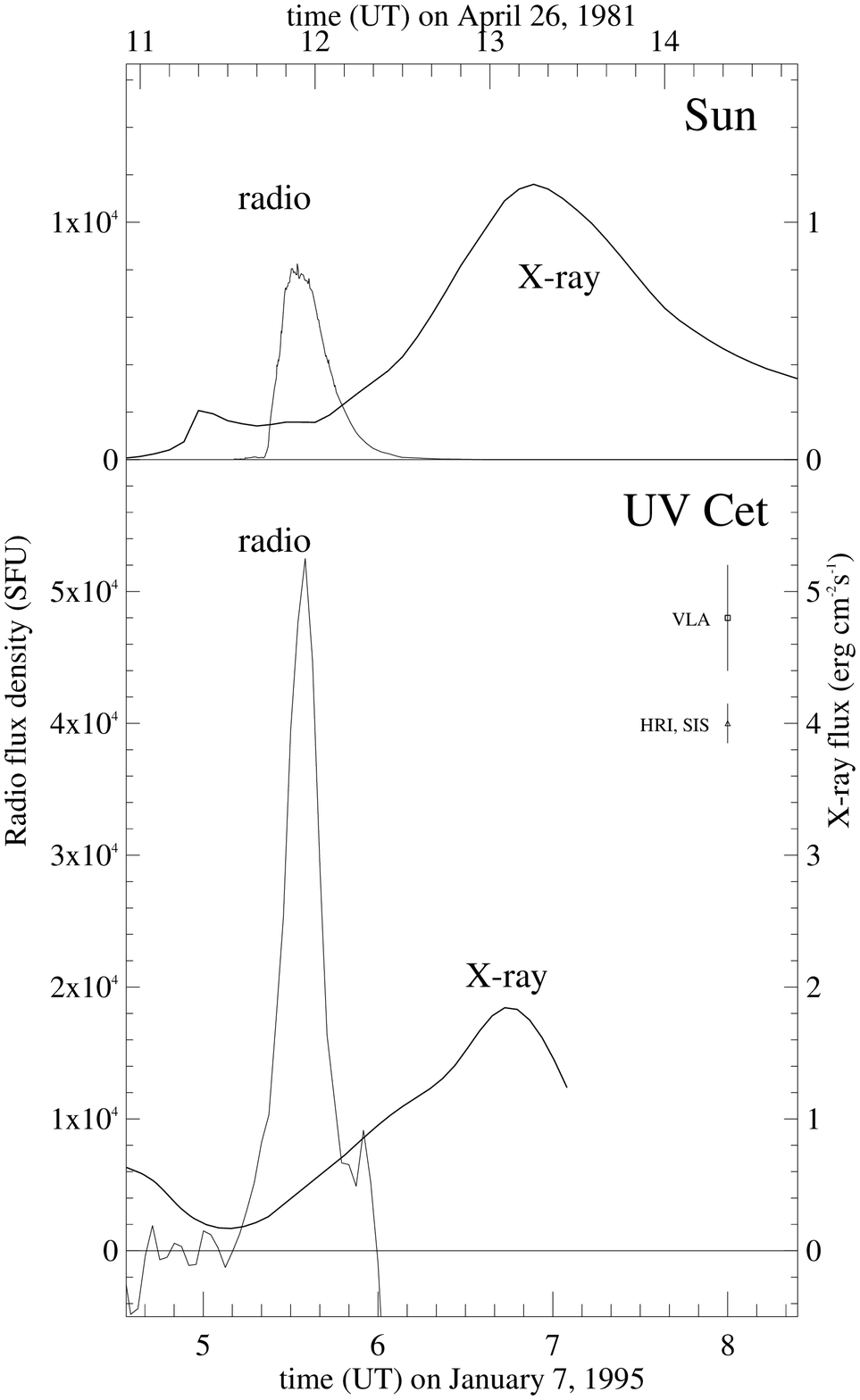}{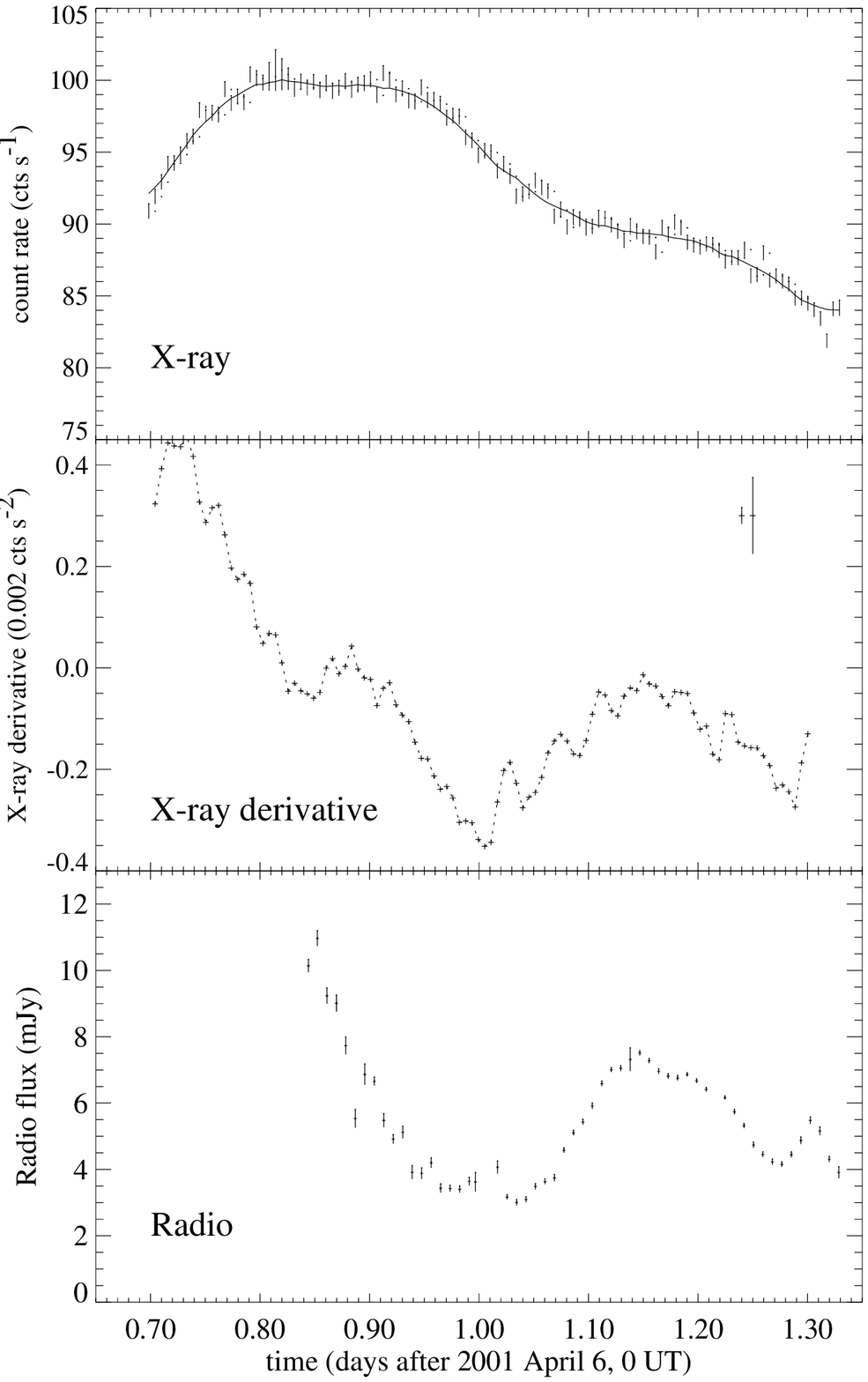}
\caption{Neupert effect seen at radio and X-ray 
        wavelengths. {\bf Left}: Lower panel shows example on the dMe star UV Cet,
	and upper panel a similar solar case (from G\"udel et al. 1996). {\bf Right:}
	Example from RS CVn binary $\sigma$ Gem. Top: X-ray light curve. Middle:
	Time derivative of X-rays. Bottom: Radio light curve (after G\"udel et al.
	2001c).}
\end{figure}

\section{Energy Release and Coronal Heating in Stars: All Made of Flares?}  
	 
The very high temperatures attained in non-flaring X-ray coronae of magnetically
active stars have often been taken as a hint for the presence of {(micro)}\-flaring.
In such a model, both the quiescent radio and the quiescent X-ray emissions could be the consequence
of many superimposed  flares, predicting that the two emissions relate to each other the same
way as they do in flares.

Figure 3  plots average loss rates in X-rays vs radio, for quiescent stars,
for {\it solar} flares, and for three flares observed on UV Cet. Evidently, flares
and quiescent emission follow the same trend (see G\"udel \& Benz 1993, Benz 
\& G\"udel 1994,  G\"udel et al. 1996). Is it conceivable that all of the 
quiescent emission is made up of flares?  
  
\begin{figure}[t!]
\plotone{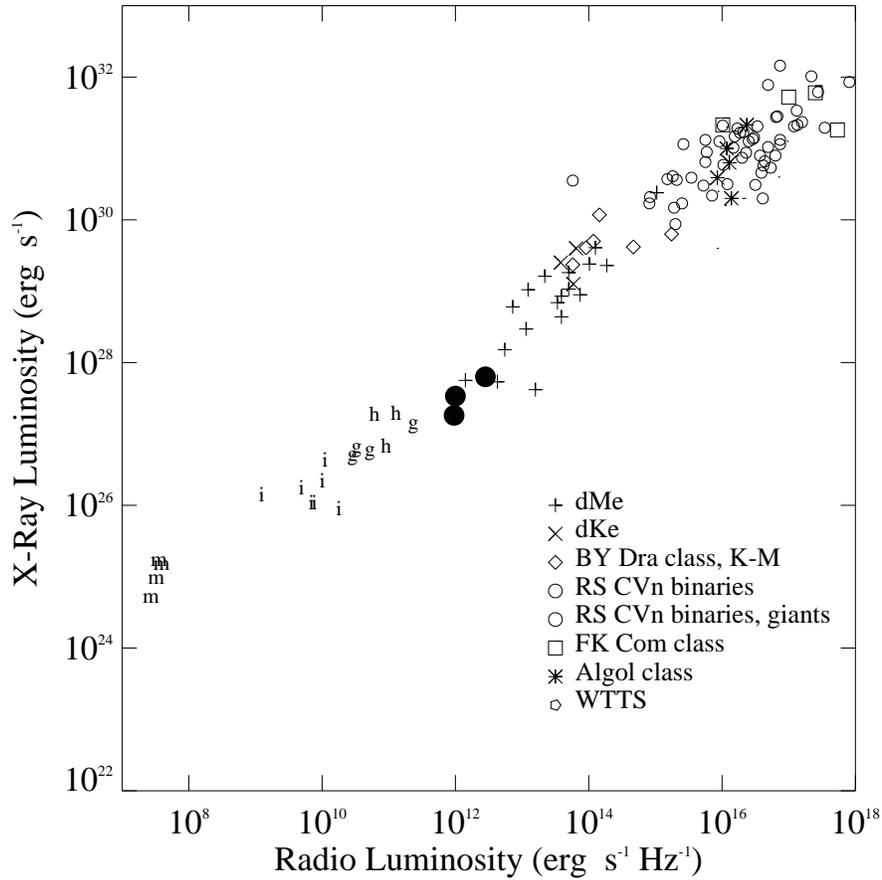}
\caption{Correlation between X-ray and radio luminosities. Upper right half of diagram
        shows quiescent luminosities of various stellar classes by different symbols.
	Lower left part shows average radiative losses during a sample of {\it solar}
	flares (letters m, i, g, h); the three solid circles mark the loci of
	three flares on UV Cet for which the Neupert effect has been observed 
	(average losses over flare durations after Benz \& G\"udel 1994 and 
	G\"udel et al. 1996).}
\end{figure}

Audard et al. (2000) statistically studied the flare energy distribution at EUV 
wavelengths to conclude the flares are distributed in energy according to a power-law,
$dN/dE \propto E^{-\alpha}$ where $\alpha > 2$ is suggested. Similar results
have been obtained by G\"udel et al. (these proceedings) and 
Kashyap  et al. (these proceedings) from a statistical investigation of a 
long monitoring of AD Leo. The following predictions should then be verified:
\begin{itemize}
\item [1)] The rate of detected flares should increase with increasing quiescent
      luminosity (this is a necessary, but not sufficient condition for the flare 
      heating hypothesis, see Audard et al. 2000);
       
\item [2)] The X-ray emission measure distribution should be compatible with that
      of a statistical distribution of rising and decaying flares;  
      
\item [3)] The radio emission should strongly be related to the {\it hot} plasma component
      of the X-ray emission measure distribution. More active stars should 
      show both more radio emission and a hot tail in the EM distribution.
\end{itemize}
There is evidence for all three predictions to hold (Audard et al. 2000 for [1],
G\"udel 1997, G\"udel, Guinan, \& Skinner 1997, and G\"udel et al.  [these proceedings] for
[2], and G\"udel et al. 1997, 1998 for [3] - see below).

\section{Particle Acceleration and Coronal Heating}  
	
If flares and non-thermal particle populations are ultimately responsible for coronal 
heating, one should see further evidence in the end-product - the hot plasma
itself. Correlating the {\it hot} ($> 10$~MK) emission measure of active
stars (as determined typically by low-resolution devices such as {\it ASCA}) with
the overall quiescent radio luminosity indeed shows a tight correlation, see Figure
4a. The large range in both parameters involved is likely to be due to a large range
in coronal volume (from coronae of solar analogs to giants), but the correlation 
can nevertheless not be reduced to a  trivial consequence of geometric effects.
It rather indicates that the presence of non-thermal particles specifically correlates
with the presence of {\it hot} plasma. Note that this correlation breaks
down for cooler, few-MK plasma: The Sun does not maintain any appreciable quiescent
emission, nor does it show quiescent plasma above 10~MK.

\begin{figure}[t!]
\plottwo{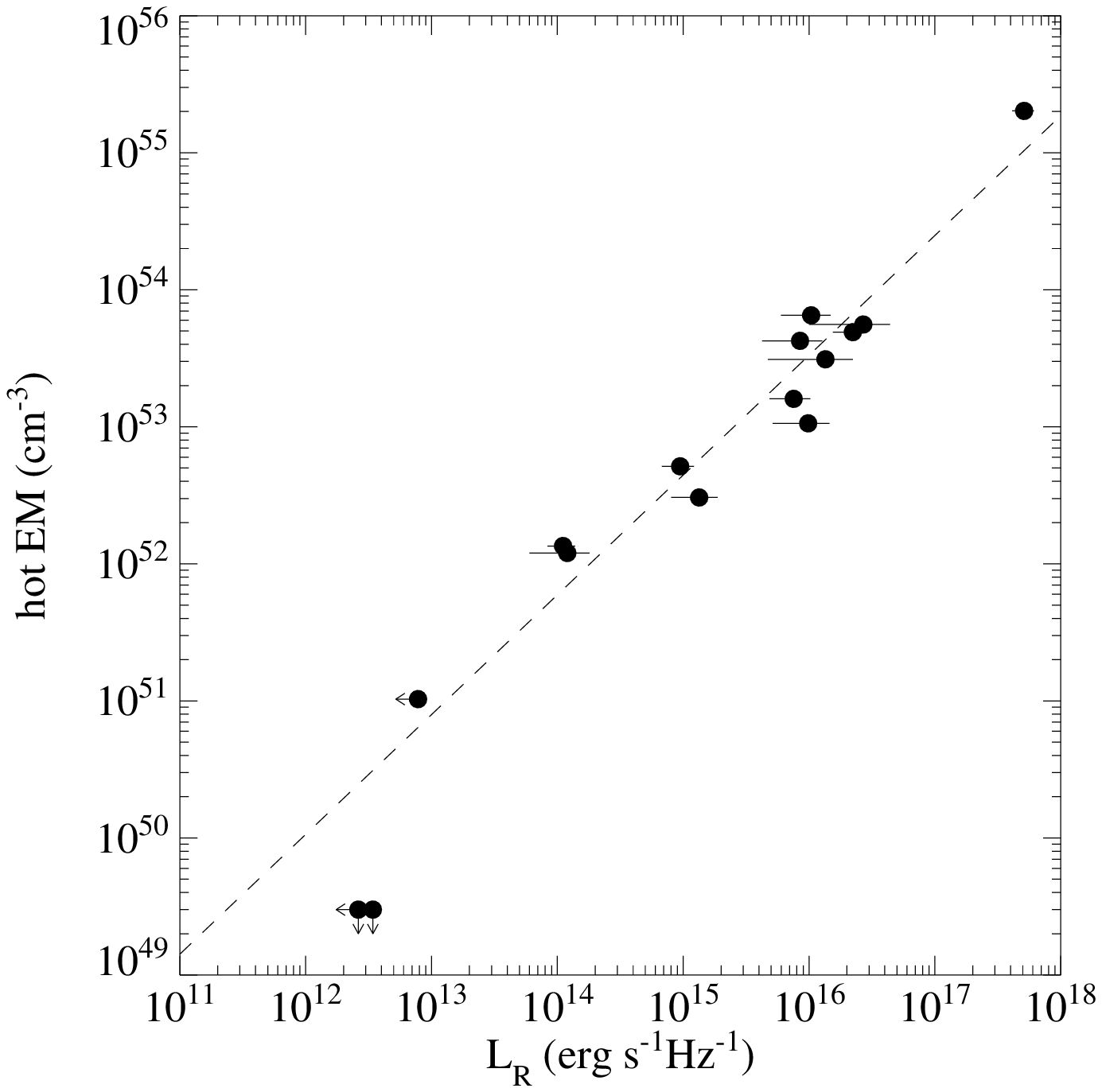}{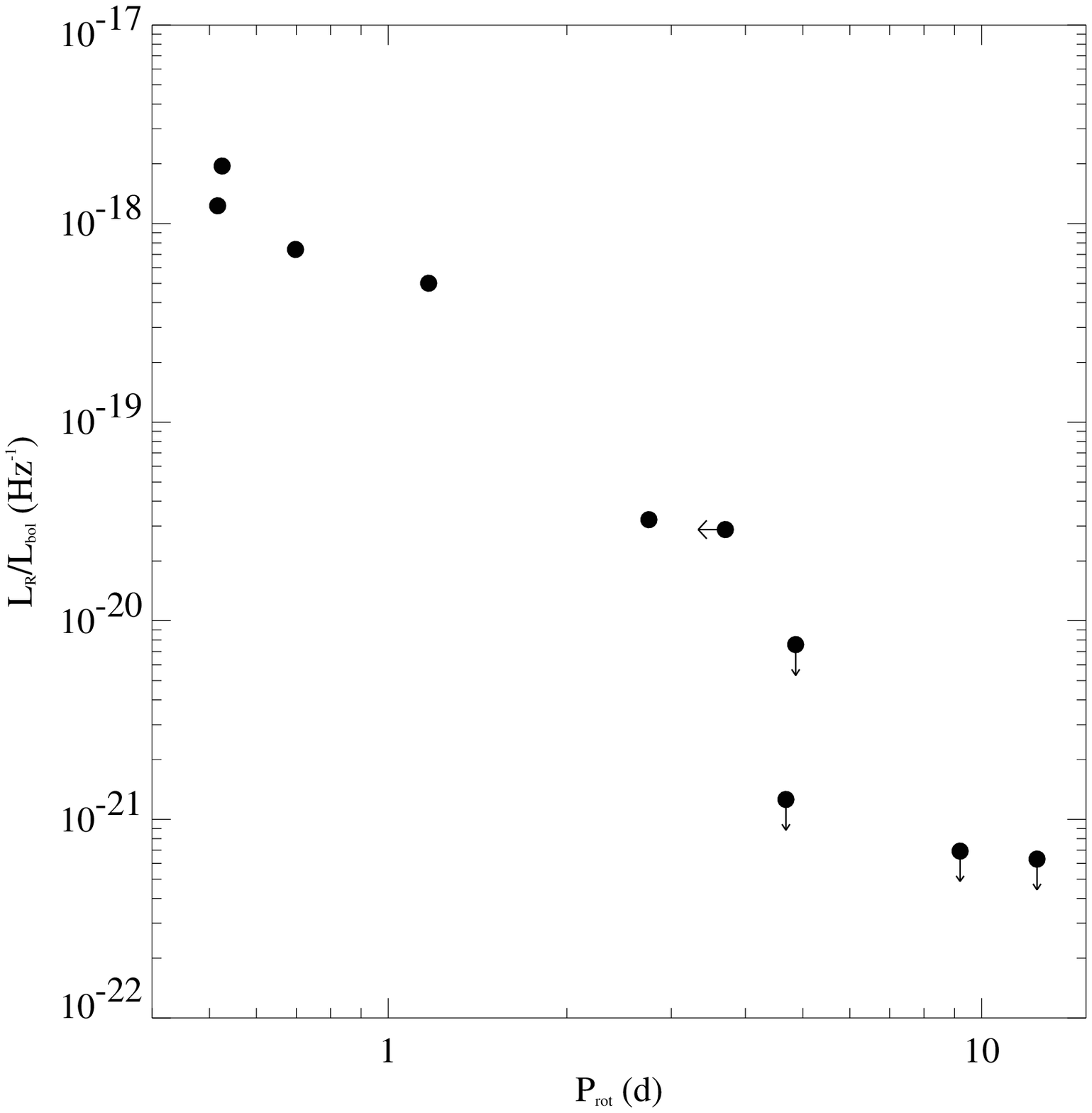} 
\caption{
	 {\bf Left:} Dependence between radio luminosity and hot 
	 coronal plasma.
         {\bf Right:} Dependence of quiescent radio emission on rotation period for
         solar analogs. (Details and references: G\"udel et al. 1998, G\"udel \& Gaidos 2001d.)
	 }
\end{figure}

The mutual dependence between accelerated particles and hot plasma is also visible as an
evolutionary effect. Since the X-ray luminosity $L_{\mathrm{X}}$ of a given class of cool 
stars correlates with the average coronal temperature (Figure 5) and $L_{\mathrm{X}}$ decays with
age,  the coronal temperature decays with age as well (G\"udel et al.
1997). The expected accompanying, rapid decay of the quiescent radio luminosity
with increasing rotation period (thus, increasing age) is shown in Figure 4b for
solar analogs. 

\begin{figure}[t!]
\plotone{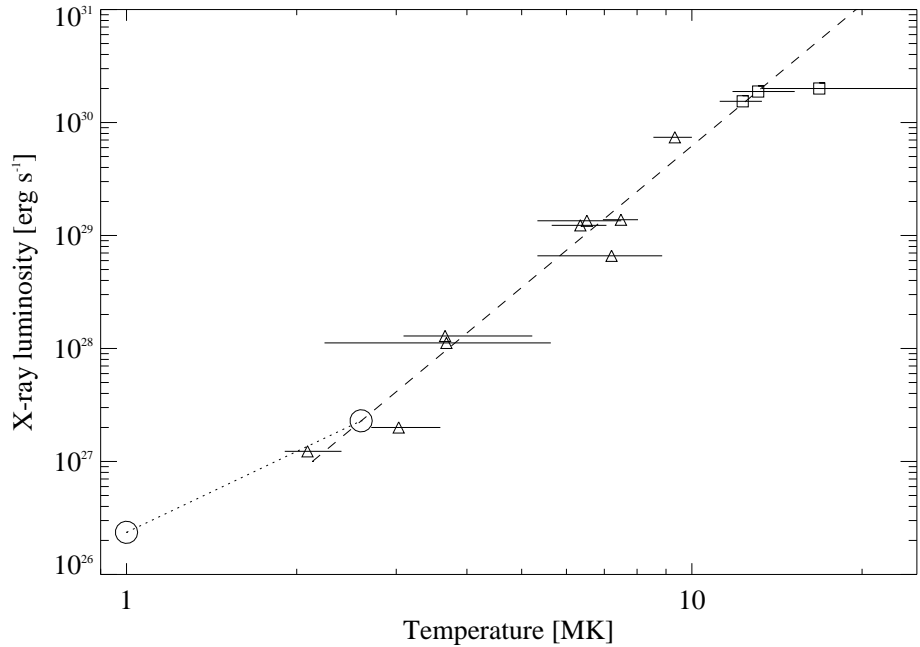} 
\caption{Dependence of average coronal temperature of solar analogs (sample from 
      G\"udel et al. 1997) on the
      overall X-ray luminosity. The circles give the range for the 
      solar corona after Peres et al. 2000.}
\end{figure}
                 
\section{Conclusions}

Stellar coronae show many facets as soon as we open our view to a variety
of wavelengths. Some high-energy processes cannot yet be studied on stars
due to sensitivity limits. However, results at hand suggest a very intimate
interplay between magnetic fields, high-energy processes of energy release,
particle acceleration, and coronal heating both during flares and during quiescence:
\begin{itemize}
\item Non-thermal coronae show a bewildering variety of sizes and structures
      that can just be resolved with VLBI techniques. No solar analogy exists.  
\item Non-thermal stellar coronae are likely to be separate entities and are
      not co-spatial with soft X-ray coronae.
\item A large amount of energy is available in accelerated particles during 
      stellar flares. They may partly be responsible for coronal heating 
      during flares via chromospheric evaporation as suggested by the presence
      of the Neupert effect.
\item There is various evidence that quiescent stellar emission (both non-thermal
      and hot-thermal) is a consequence of superimposed flaring
      events.
\end{itemize}
Evidently, the study of the interplay between high-energy electrons and coronal
heating is important. It may link quiescent coronae to flare physics, and 
may thus relate back to solar physics where detailed study of physical mechanisms is 
easier. A notable missing link are {\it hard X-rays} from stars. There is hope that 
future large-area devices such as {\it Integral} will remedy this situation.                        

\acknowledgments
Research at PSI  has been supported by the Swiss National 
Science Foundation (grant 2100-049343).  The VLA is a facility of the 
National Radio Astronomy 
Observatory, which is operated by Associated Universities,
Inc., under cooperative agreement with the  National Science Foundation.

\end{document}